\documentclass[%
 aip,
 jcp,
 amsmath,amssymb,
reprint, 
]{revtex4-1}

\usepackage{graphicx}
\usepackage{dcolumn}
\usepackage{bm}

\usepackage[utf8]{inputenc}
\usepackage[T1]{fontenc}
\usepackage{mathptmx}
\usepackage{etoolbox}
\usepackage{hyperref}

\usepackage{xcolor}
\definecolor{vbred}{rgb}{0.0, 0.0, 0.0}
\newcommand{\review}[1]{{\color{vbred}{#1}}}

\hypersetup{
    colorlinks=true,
    linkcolor=blue,
    filecolor=magenta,      
    urlcolor=blue,
    pdfpagemode=FullScreen,
    citecolor=blue,
    }

\makeatletter
\def\@email#1#2{%
 \endgroup
 \patchcmd{\titleblock@produce}
  {\frontmatter@RRAPformat}
  {\frontmatter@RRAPformat{\produce@RRAP{*#1\href{mailto:#2}{#2}}}\frontmatter@RRAPformat}
  {}{}
}%
\makeatother
\begin{document}

\title{Efficient Computation of the Long-Range Exact Exchange using an Extended Screening Function}
\author{Sebastian Kokott}
\affiliation{The NOMAD Laboratory at the Fritz Haber Institute of the Max Planck Society, Berlin, Germany}%
\affiliation{Molecular Simulations from First Principles e.V., Berlin, Germany}
\email{kokott@fhi-berlin.mpg.de}
\author{Volker Blum}
\affiliation{Thomas Lord Department of Mechanical Engineering and Material Science, Duke University, Durham, North Carolina, USA}
\affiliation{Department of Chemistry, Duke University, Durham, North Carolina, USA}

\author{Matthias Scheffler}
\affiliation{The NOMAD Laboratory at the Fritz Haber Institute of the Max Planck Society, Berlin, Germany}

\begin{abstract}
We introduce a computationally efficient screening for the Coulomb potential that also allows calculating approximated long-range exact exchange contributions with an accuracy similar to an explicit full-range evaluation of the exact exchange.
Starting from the screening function of the HSE functional, i.e., the complementary error function, as zeroth order, a first-order Taylor expansion in terms of the screening parameter $\omega$ is proposed as an approximation of the long-range Coulomb potential. The resulting extended screening function has a similar spatial extend as the complementary error function leading to a computational speed comparable to screened hybrid functionals such as HSE06, but with long-range exact exchange contributions included. 
The approach is tested and demonstrated for prototypical semiconductors and organic crystals using the PBE0 functional. Predicted energy band gaps, total energies, cohesive energies, and lattice energies from the first-order approximated PBE0 functional are close to those from the unmodified PBE0 functional, but are obtained at significantly reduced computational cost.
\end{abstract}

\maketitle

\section{Introduction}

Hybrid density functional approximations (DFAs) often stand out for their accuracy in performing ab initio electronic-structure based simulations compared to more limited semilocal DFAs. In general, hybrid density functionals have a rigorous foundation in the framework of generalized Kohn-Sham (gKS) density functional theory (DFT)~\cite{seidl1996generalized,gorling1997hybrid,garrick2020exact}. The gKS band gap coincides with the fundamental band gap as defined by total-energy differences for solids~\cite{perdew2017understanding}. Thus, hybrid DFAs have the potential to predict reasonable values for band gaps in principle.
In practice, hybrid DFAs offer improvements over generalized gradients approximations (GGAs) and meta-GGA functionals for several electronic properties, especially when charge transfer, energy barriers, or localisation play an important role~\cite{goerigk2010,gao2013electronic,Tran2020}.
Moreoever, linear scaling hybrid DFA implementations have been achieved for several approaches~\cite{Schwegler1997,Ochsenfeld1998,ihrig2015,Bussy2024,forster2020double,hu2017interpolative,Dziedzic2013,wu2009order,ko2020enabling,lin2016adaptively,levchenko2015} and, thus, hybrid DFAs are also suited for large-scale density functional simulations. 

The highly successful HSE screened exchange functional~\cite{heyd2003,heyd2006} was originally introduced in order to enable simulations of molecules and periodic solids with the accuracy of a hybrid DFA while limiting the computational cost compared to functionals with full-range exact exchange. The HSE functional uses the complementary error function erfc as a screening function, which limits the Coulomb potential to a finite range. In turn, the numerical evaluation of exact exchange is accelerated by not computing the long-range contributions.~\cite{moussa2012analysis} As recently demonstrated, screened hybrids can be used for all-electron {\it ab initio} simulations up to 30,000 atoms \cite{kokott2024efficient}. At the same time, the erfc function also mimics the electronic screening in solids, which explains the qualitative improvements of certain computational predictions. For example, the HSE06 functional outperformed several other hybrid DFAs, when tested across a large benchmark of inorganic solids for the prediction of the band gap.~\cite{borlido2019large}. Nevertheless, for some material classes the long-range exact exchange contribution is of significant importance. For organic materials, functionals with long-range exact exchange contributions appear to offer a notable accuracy advantage for predictions of the structure of organic crystals, and therefore often emerge as a preferred reference methods for such studies~\cite{hoja2019reliable,neumann2008major,price2023xdm}. Moreover, in materials with a band gap the Coulomb potential should decay asymptotically as $1/(\epsilon_\infty r)$~\cite{cappellini1993model,Marques2011}, where $\epsilon_\infty$ refers to the static dielectric constant. In turn, also inorganic materials should have long-range exact exchange contributions. Several approaches explore this fact by either using a system dependent mixing of the fraction of exact exchange in the hybrid functional~\cite{Marques2011,skone2014self,khan2024adaptive}, or by mixing in long-range exact exchange contributions separately (range-separated hybrids, e.g. see Ref.~\onlinecite{baer2010tuned}), or model screening functions are used as a range-dependent mixing parameter (range-separated dielectric-dependent hybrid functionals~\cite{skone2016nonempirical,chen2018nonempirical} or screened exchange functionals~\cite{lorke2020koopmans}).
Unfortunately, the computational evaluation of the long-range EXX part remains demanding, as the Coulomb tail contributes notably to the cost of simulating periodic solids.

Hybrid DFAs mix the (semi)local exchange of the local density approximation (LDA), GGAs, or meta-GGAs with some fraction of non-local exact exchange (EXX). In general, most hybrid DFAs can be expressed within the following scheme (same notation as used in the exchange-correlation functional library libxc~\cite{lehtola2018recent}): 
\begin{align}
    E_x(\alpha,\beta,\omega) &= \alpha  E_\text{exx} + \beta E^\text{SR}_\text{exx}(\omega) \nonumber \\ &+ (1-\alpha) E_\text{x-DFA} - \beta E^\text{SR}_\text{x-DFA}(\omega),
    \label{eq:global_hybrids}
\end{align}
with the exact exchange energy $E_\text{exx}$, the short-range exchange energy $E^\text{SR}_\text{exx}(\omega)$, the full-range DFA exchange energy $E_\text{x-DFA}$, and the short-range DFA exchange energy $E^\text{SR}_\text{x-DFA}(\omega)$. For the division into long- and short-range part different choices exist, e.g., the error function or the Yukawa screening. In this work, the range-separation always uses the error function. Using the notation in Eq.~\eqref{eq:global_hybrids}, e.g., the PBE0 functional~\cite{ernzerhof1999,adamo1999} can be obtained by $\alpha=0.25$ and $\beta=0$, the HSE06 functional~\cite{heyd2003,heyd2006} by setting $\alpha=0.0$, $\beta=0.25$ , and $\omega=0.11$ Bohr$^{-1}$, and the $\omega$B97 functional by using $\alpha=1$, $\beta=-1$, and $\omega=0.4$ Bohr$^{-1}$ and using B97 screened exchange~\cite{chai2008systematic} for the DFA contribution. Hybrid DFAs with $\beta=0$ are referred to as global hybrid DFAs and hybrid DFAs with $\beta>0$ as range-separated hybrids. The parameters $\alpha$ and $\alpha + \beta$ describe the long-range and short-range exact exchange contribution, respectively. Hybrid DFAs with $\alpha =0$, $\beta,\omega>0$, like the HSE functional, only compute the short-range part of the exact exchange. Hybrids with $\alpha>0$ compute both short-range and long-range exact exchange.

In this paper, we derive an extension of the HSE screening function erfc that includes an approximation of the long-range Coulomb potential within a (controllable) finite range. The approximation is obtained by a first-order Taylor expansion starting from the HSE screening function erfc. The final extended screening function has a localization similar to the conventionally used erfc function. As derived, the shape of our extended screening function is not motivated by an actual physical screening mechanism for the electrons. Rather, if this screening function is used for the exact exchange contribution, the evaluation of the exact exchange term remains computationally efficient, similar to the HSE functional, but it also includes long-range exact exchange contributions, similarly to when the unscreenened Coulomb potential is used. We demonstrate this approach for the PBE0 functional. We will refer to the original PBE0 functional as "unmodified PBE0" and the functional using our proposed screening function as "first-order approximated PBE0" or PBE0$^\prime$ throughout the paper. We note that this approach is not restricted to the PBE0 functional, but can be used for all hybrid functionals that include the long-range EXX contributions, e.g. functionals like B3LYP~\cite{stephens1994}, $\omega$B97, or M11~\cite{peverati2011}. 

\section{Approximating the long-range Coulomb potential using a screening function}

\begin{figure*}[t]
    \centering
    \includegraphics[width=\textwidth]{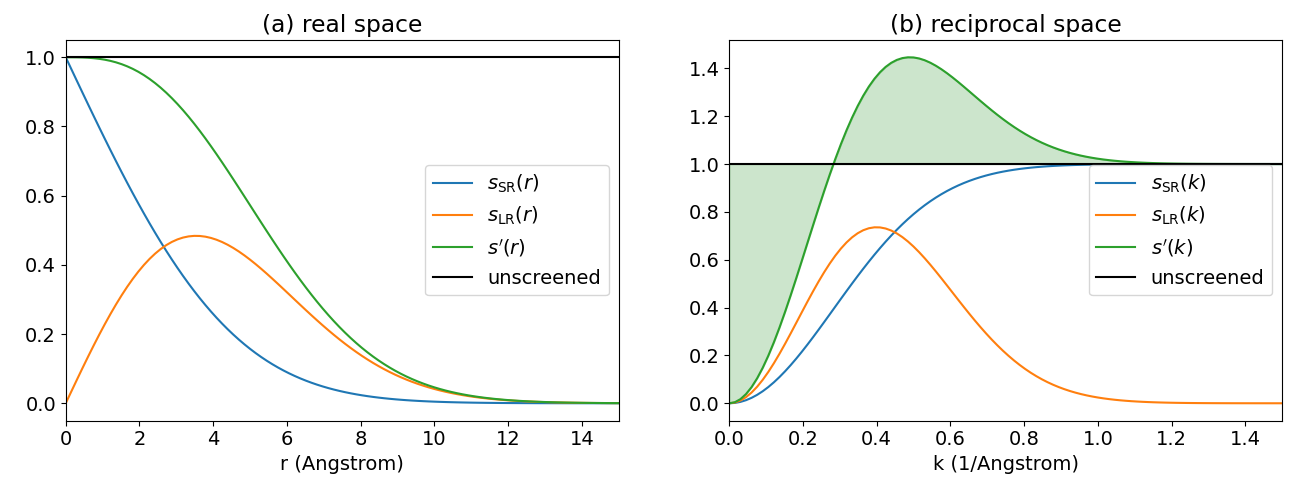}
    \caption{(a) The short-range screening function $s_\text{SR}$ (Eq.~\eqref{s_SR}; blue line), the derived long-range screening function $s_\text{LR}$ (Eq.~\eqref{s_LR}; orange line), and our proposed screening function (Eq.~\eqref{s_prime}; green line) as a function of the distance $r$. (b) The corresponding Fourier transforms of the screening functions for the short-range part (blue line), the long-range part (orange line), and the one with both parts combined (green line) in reciprocal space. The two green areas in (b) are of same size. The range-separation parameter $\omega$ is set to 0.11~\text{Bohr}$^{-1}$.}
    \label{fig:range_separation}
\end{figure*}

For hybrid density functionals, the evaluation of the exchange operator is often the dominant step in the simulation time. The approach outlined in this work will work best for evaluations of the exact exchange in a real-space formalism. In real space, the evaluation time of $X(\mathbf{R})$ can be significantly reduced by a screening function, such as the erfc function, because the sparsity of the exchange matrix $X(\mathbf{R})$ increases. The exchange operator in real-space can be formulated in the following way:
\begin{widetext}
\begin{align}
    X^\sigma_{ij} (\mathbf{R}) = \sum_{kl}\sum_{R^\prime} D^\sigma_{kl}(\mathbf{R}^\prime) \sum_{R^{\prime\prime}}\iint\text{d}\mathbf{r}\text{d}\mathbf{r^\prime} \varphi_i(\mathbf{r}) \varphi_k(\mathbf{r} - \mathbf{R}^{\prime\prime}) \varphi_j(\mathbf{r}^\prime - \mathbf{R}) \varphi_l(\mathbf{r}^\prime - \mathbf{R}^\prime - \mathbf{R}^{\prime\prime}) V(\mathbf{r} - \mathbf{r}^\prime)
    \label{eq:exchange_operator}
\end{align}
\end{widetext}
with the real-space lattice vector within the Born-von Karman cell $\mathbf{R}$, the spin density index $\sigma$, density matrix $D^\sigma_{kl}$. The indices $i$, $j$, $k$, $l$ denote the basis functions $\varphi$ localized at the corresponding atoms. The Coulomb kernel $V(\mathbf{r} - \mathbf{r}^\prime)$ refers to any functional dependent weighting of the short-range (SR) and long-range (LR) contribution of the Coulomb potential $v$:
\begin{align}
    v(r) = 1/r=  v_\text{SR}(r;\omega) + v_\text{LR}(r;\omega) \nonumber \\ = \frac{\text{erfc}(\omega r)}{r} + \frac{\text{erf}(\omega r)}{r},
    \label{eq:hse06}
\end{align}
where erf (erfc) refers to the (complementary) error function. \review{To achieve an efficient evaluation of Eq.~\eqref{eq:hse06}, we use the localized resolution-of-identity approach RI-LVL~\cite{watson2003density,ihrig2015,wang2020efficient}, described with full technical details for FHI-aims in Ref.~\cite{ihrig2015}.

For $\omega \rightarrow 0$ the long-range contribution becomes zero and the short-range part $v_\text{SR}(r;\omega)$ recovers the full Coulomb potential $1/r$. In order to approximate the full Coulomb potential, we expand the short-range potential $v_\text{SR}(r;\omega)$ around $\omega$ by a Taylor series up to first order:
\begin{align}
    v_1(r;\omega^\prime) & := v_\text{SR}(r;\omega)  + (\omega^\prime -\omega) \frac{\partial v_\text{SR}(r;\omega)}{\partial \omega} + \dots 
\end{align}
The recommended value of Ref.~\onlinecite{Krukau2006} for $\omega$ in the screening function of HSE06 is 0.11 Bohr$^{-1}$. Setting $\omega^\prime=0$, $v_1(r;0)$ approximates the full Coulomb potential:
\begin{align}
    v(r) \approx v_1(r;0) = v_\text{SR}(r;\omega) -\omega \frac{\partial v_\text{SR}(r;\omega)}{\partial \omega} 
    \label{eq:approx_coulomb}
\end{align}
By comparing Eqs.~\eqref{eq:hse06} and~\eqref{eq:approx_coulomb}, the term $-\omega \frac{\partial v_\text{SR}(r;\omega)}{\partial \omega}$ is identified as an approximation to the long-range contribution $v_\text{LR}$.} Evaluating the first derivative, using erfc as the form of the screening function as, e.g. in HSE06, leads to our proposed long-range Coulomb approximation:
\begin{align}
    v_\text{LR}(r,\omega) \approx -\omega \frac{\partial v_\text{SR}(r;\omega)}{\partial \omega} = \frac{2\omega}{\sqrt{\pi}} e^{-(\omega r)^2}
    \label{eq:lr_approx}
\end{align}
The corresponding screening functions (generally denoted as s) can be obtained by multiplying the range-separated Coulomb potentials with~$r$:
\begin{align}
    s^\prime (r) &= s_\text{SR} + s_\text{LR} \label{s_prime}\\
    s_\text{SR} (r) &= \text{erfc}(\omega r) \label{s_SR}\\
    s_\text{LR} (r) &= \frac{2\omega r}{\sqrt{\pi}}e^{-(\omega r)^2}, \label{s_LR} 
\end{align}
and are shown in Figure~\ref{fig:range_separation}. Interestingly, the screening function that also governs the long-range Coulomb potential, $s^\prime$, remains localized in real space. Figure~\ref{fig:range_separation}(a) shows the original erfc function, $s_\text{SR}$, as used in the HSE06 functional (blue line) and our derived screening function $s^\prime$ (green line), which includes the long-range approximation $s_\text{LR}$ (orange line). We find that the additional contribution only marginally increases the range of the erfc function. The Fourier transformed Coulomb potential with the screening function $s^\prime$ is given by:
\begin{align}
    v(k)\approx& \mathcal{F}\left[ v_\text{SR}(r;\omega) \right] + \mathcal{F}\left[ v_\text{LR}(r;\omega) \right] \nonumber \\ =& \frac{4\pi}{k^2}(1-e^{-\frac{k^2}{4\omega^2}}) + \frac{2\pi}{\omega^2}e^{-\frac{k^2}{4\omega^2}}. \label{eq:PBE0_approx_k}
\end{align}
The corresponding screening functions in k-space are obtained by dividing Eq.~\eqref{eq:PBE0_approx_k} by the Fourier-transformed Coulomb potential $v(k) = 4\pi/k^2$ and are shown in Fig.~\ref{fig:range_separation}(b). The shown Fourier components provide a helpful interpretation on how the screening function $s^\prime$ accounts for the missing long-range contribution: The missing exact exchange part cut out by the erfc function is compensated by an increased contribution for k-values between 0.25 and 1.25 \AA$^{-1}$ as indicated by the green area above the horizontal line at 1. The other area drawn in Fig.~\ref{fig:range_separation}(b) below the horizontal line at 1 is of the same size, so that in total the weighting of the EXX contributions sum up to 1, i.e., the unscreened Coulomb potential.

We note that the long-range exact exchange contribution can be also recovered with the HSE range-separation function erfc for $\omega \rightarrow 0$. However, the convergence of this approach is much slower and it is accompanied by a significant increase of computation time and memory compared to our proposed screening function $s^\prime$ with finite $\omega$. Our model extrapolates the long-range contribution from an intermediate yet still localized range, so the functional converges much faster (as a function of $\omega$) towards the exact long-range EXX contribution than the screened erfc potential (the zeroth order in Eq.~\eqref{eq:approx_coulomb} on its own. The importance of the intermediate range for the exact exchange contributions has been analyzed before in Ref.~\onlinecite{henderson2007importance,lucero2012improved}. 

\section{Results and Discussion}

In the following section, we validate our screening function $s^\prime$ for the PBE0 functional with the standard mixing parameter $\alpha=0.25$. The purpose of the fixed mixing parameter is to systematically study the effect of our proposed screening function $s^\prime$ across all materials. As noted earlier, we  refer to the PBE0 functional using the screening function $s^\prime$ in the EXX contribution as PBE0$^\prime$. \review{The final working equation for the exchange energy $E_\text{x}$ of the PBE0$^\prime$ is given by: 
\begin{align}
    E_\text{x}(\omega) = \frac{1}{4} E_\text{EXX}^\prime(\omega) + \frac{3}{4} E_\text{PBE-X},
\end{align}
where $E_\text{EXX}^\prime(\omega)$ refers to the exact exchange energy with screening function $s^\prime$ (Eq.~\eqref{s_prime}) in the exchange operator Eq.~\eqref{eq:exchange_operator}, with range-separation parameter $\omega$, and $E_\text{PBE-X}$ refers to the PBE exchange~\cite{perdew1996generalized}.} We use the unmodified PBE0 functional as reference to measure the deviations in the PBE0$^\prime$ functional. For the screening function $s^\prime$, we use the range separation parameter $\omega = 0.11$~Bohr$^{-1}$ if not noted otherwise. The benchmark consists of 16 inorganic and 23 organic crystals. All calculations are carried out with the all-electron code FHI-aims~\cite{blum2009} and the real-space localized resolution-of-identity approach RI-LVL~\cite{ihrig2015,levchenko2015,kokott2024efficient} for the EXX contributions. The RI-LVL approach in FHI-aims facilitates the sparsity by compression of several matrices. As described before, a screening function that effectively limits the range of the Coulomb potential increases the sparsity of matrices and, thus, allows for a faster evaluation of the EXX contribution. \review{We use the {\it intermediate} species defaults~\cite{blum2025} of FHI-aims (as shipped with the FHI-aims source code) for the NAO basis set.} \review{We use a k-grid density that corresponds to the size of a Born-von Karman cell of at least $40~\text{\AA} > n_i \cdot a_i $, where $a_i$ is the lattice vector length and  $n_i$ is the number of k-points along the corresponding k-space direction~$i$. This  size of the Born-von Karman cell ensures that all non-zero exact exchange contributions due to the overlap of the NAO basis functions are included. One important consequence of this choice is that the Spencer-Alavi~\cite{spencer2008efficient} truncation of the exact exchange term to the Born-von Karman cell, which is implemented in FHI-aims~\cite{levchenko2015}, has no effect for either the PBE0 functional or for the PBE0$^\prime$ functional} 

\subsection{Inorganic crystals}

\begin{figure*}[t]
    \centering
    \includegraphics[width=\textwidth]{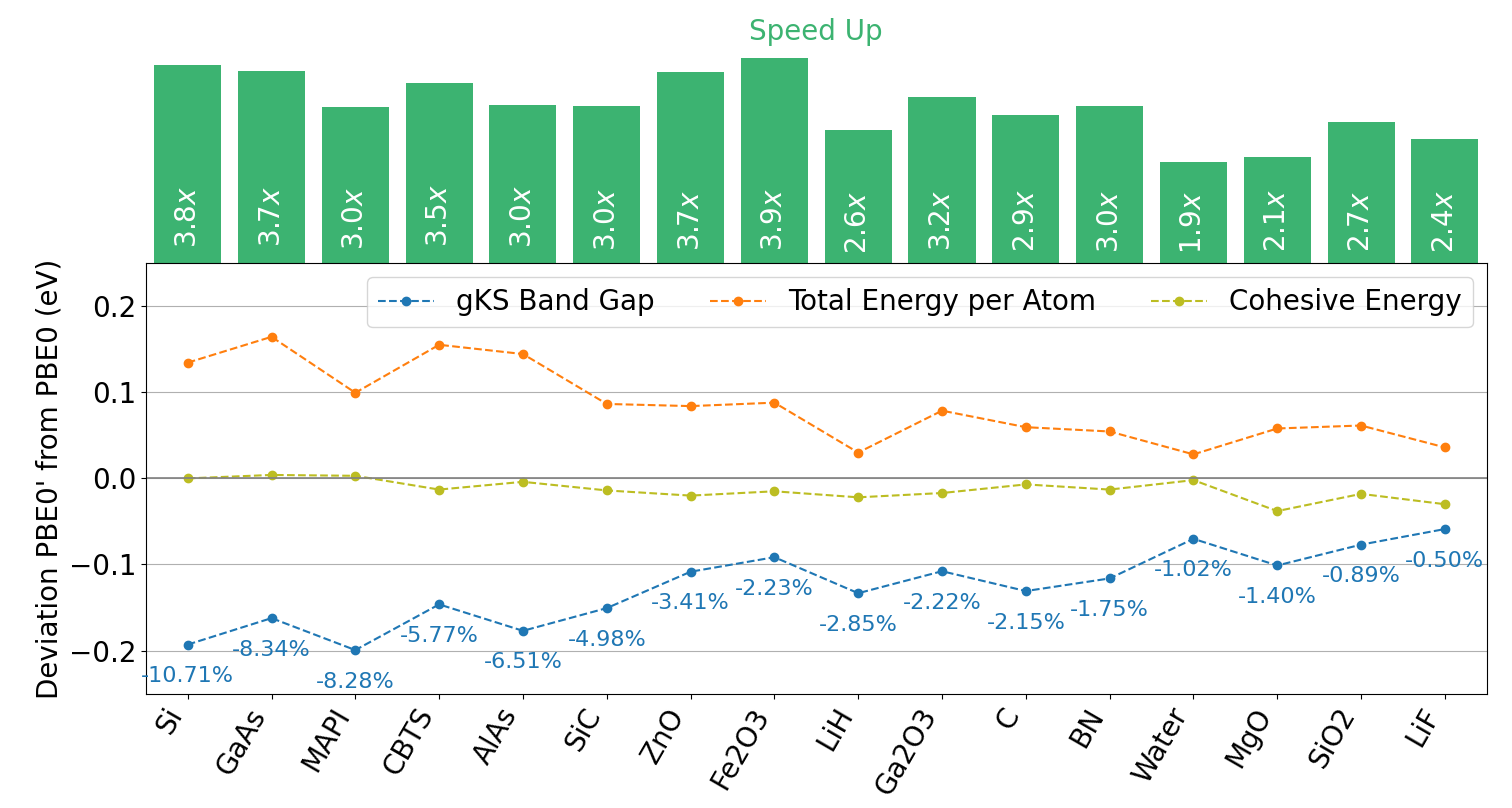}
    \caption{Deviations of the PBE0$^\prime$ functional from the PBE0 functional for the generalized Kohn-Sham (gKS) band gap (blue line), the total energy per atom (orange line), and the cohesive energy per atom (olive line) for 16 different inorganic materials. The materials are sorted from left to right with increasing PBE0 gKS band gap size. The green bars at the top give the speed-up for each material compared to the unmodified PBE0 functional. The annotated numbers for the gKS band gaps are the relative errors w.r.t. to the PBE0 gKS band gaps. All calculations were carried out on the Raven HPC cluster at the MPCDF using Intel Xeon IceLake (Platinum 8360Y) nodes with 72 cores per node.}
    \label{fig:inorganics}
\end{figure*}

The set of 16 inorganic crystals contains intermediate to wide band gap materials. A detailed listing is provided in the Supplemental Information, Table~1, defining stoichiometry, space group and abbreviations used for each system. Results for computational speedups of the simulation time for PBE0$^\prime$ (start to finish; compared to the evaluation of the unmodified PBE0 functional) and for the energy band gaps predicted for the inorganic crystal benchmark set is given in Figure~\ref{fig:inorganics} and the corresponding absolute numerical results and additional information are given in the Supplemental Information, Table~3. We observe remarkable speed-up factors, e.g. up to 3.9 for PBE0$^\prime$ for Fe$_2$O$_3$, when compared to the evaluation of the unmodified PBE0 functional. However, at the same time we observe the trend that the deviations in the gKS band gap and the total energy correlate with the band gap size: the larger the gKS band gap, the smaller the deviation of the  PBE0$^\prime$ from the PBE0 functional. For the PBE0$^\prime$ functional, we find a maximum deviation from the PBE0 gKS band gap of -0.2 eV for silicon and methylammonium lead iodide (MAPI). The negative sign of a deviation in Fig.~\ref{fig:inorganics} indicates that the PBE0$^\prime$ energy is smaller than the PBE0 energy. At the same time, we also find that the speed-up factors correlate with the band gap size: the larger the band gap, the smaller the speed up. This is consistent with the underlying physical correlation between gap size and electron localization around atoms and molecules. Usually, the larger the gKS band gap, the more localized are the valence electrons and, in turn, the sparser the density matrix. The sparsity of the density matrix $D(\mathbf{R})$ in real-space impacts the sparsity patterns in the EXX matrix $X(\mathbf{R})$ to a large extent as can be seen from Eq.~\eqref{eq:exchange_operator}. We also computed the equilibrium lattice constant $a_0$ and the bulk modulus $B$ for the cubic systems in the benchmark, given in Table~\ref{tab:lattice_bulkmodulus}. Here, the PBE0$^\prime$ functional reproduces the unmodified PBE0 results almost perfectly indicating that the potential energy surface close to the equilibrium remains intact and that errors in the total energy are a constant offset and do not affect forces and stress. This is supported by the small errors in the cohesive energy per atom shown in Fig.~\ref{fig:inorganics}. The deviation in the cohesive energies remains close to zero with a maximum deviation of 0.03 eV for MgO. It seems that the error in the PBE0$^\prime$ total energy mainly originates from the on-site contributions, which is why the difference of the total energies of the free atoms contains a large portion of the error.
We also computed the runtimes for the HSE06 functional for two examples, so we can compare them with PBE0$^\prime$ and PBE0. For Si, we obtain a total runtime of 1.45 CPU hours for HSE06, 2.02 CPU hours for PBE0$^\prime$, and 7.71 CPU hours for PBE0. Similarly for 64 water molecules, we obtain a total runtime of 23.75 CPU hours for HSE06, 28.20 CPU hours for PBE0$^\prime$, and 54.82 CPU hours for PBE0. Clearly, the runtimes for the PBE0$^\prime$ are still slower than HSE06, but much lower than PBE0 and the overhead in evaluation time accounts for long-range exact exchange contributions that are comparable as in PBE0. In short, PBE0$^\prime$ delivers almost PBE0 results at almost the computational cost of HSE06.

\begin{figure*}[t]
    \centering
    \includegraphics[width=\textwidth]{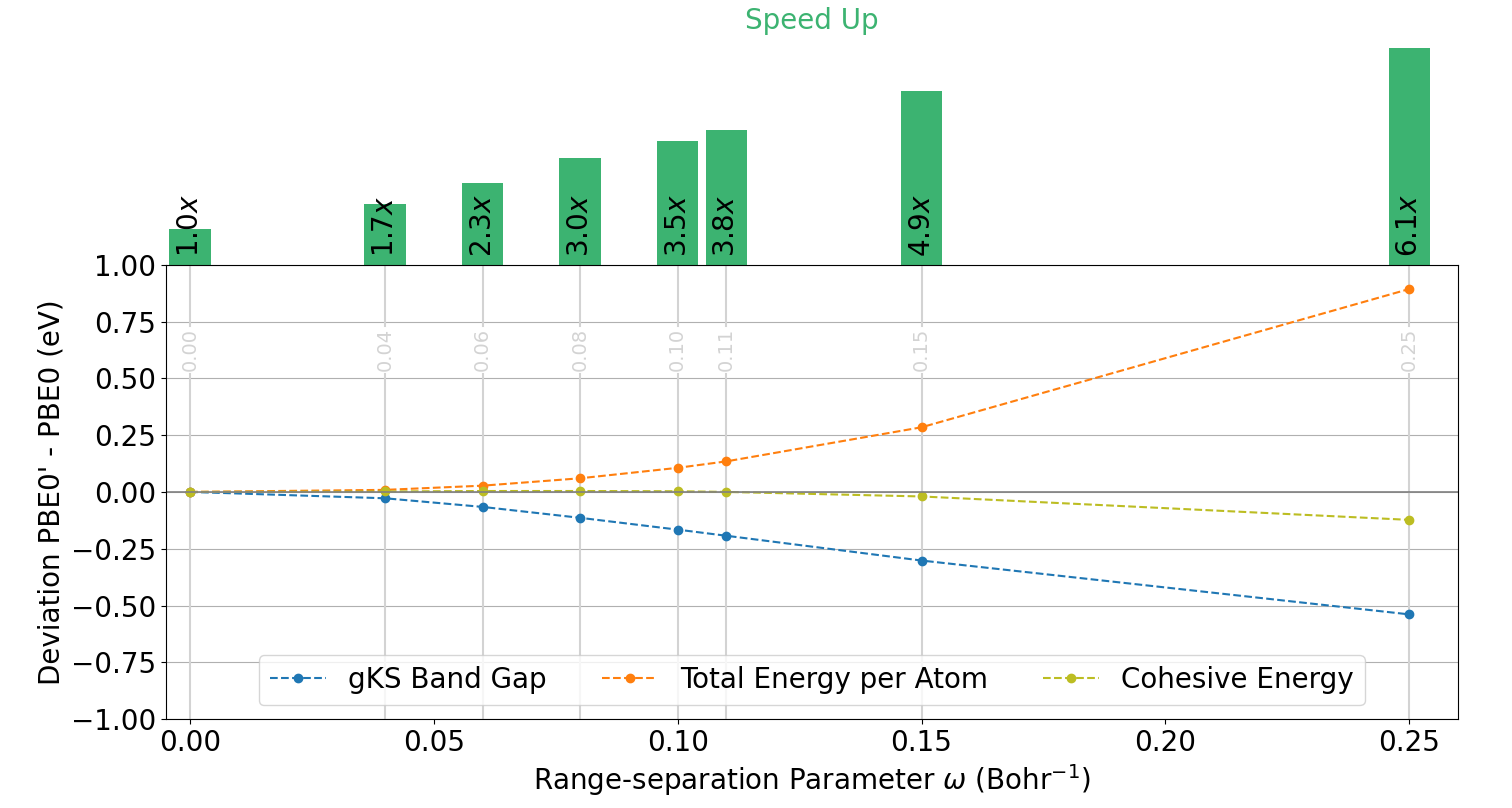}
    \caption{Deviations of the PBE0$^\prime$ functional from the PBE0 functional for the generalized Kohn-Sham (gKS) band gap (blue line), the total energy per atom (orange line), and the cohesive energy per atom (olive line) as a function of the range-separation parameter $\omega$ for Si diamond.}
    \label{fig:errors}
\end{figure*}

\begin{table}[ht]
    \centering
    \begin{tabular}{l|r|r}
        Name & $a_0$ (\AA)    & $B$ (GPa) \\ \hline
        AlAs & 5.673 (-0.005) &  77.3 (+1.4) \\
        BN   & 3.596 (-0.002) & 405.6 (+2.3) \\
        C    & 3.543 (-0.002) & 475.1 (+2.8) \\
        GaAs & 5.668 (-0.010) &  74.7 (+1.6) \\
        LiF  & 4.011 (-0.003) &  74.9 (-0.1) \\
        LiH  & 3.989 (-0.005) &  40.4 (+1.4) \\
        MgO  & 4.206 (-0.003) & 167.4 (+1.7) \\
        Si   & 5.437 (-0.007) & 100.3 (+1.7) \\
        SiC  & 5.487 (-0.005) & 234.2 (+2.4) \\
    \end{tabular}
    \caption{Lattice constants $a_0$ and Bulk moduli $B$ computed with the PBE0$^\prime$ functional for the cubic systems from Tab.~1 in the Supplemental Information. The deviation to the unmodified PBE0 are given in parentheses.}
    \label{tab:lattice_bulkmodulus}
\end{table}

As a next step, we demonstrate the impact of the range-separation parameter $\omega$ in the scope of our range-separation function $s^\prime$. In Figure~\ref{fig:errors}, we varied the rang-separation parameter from 0 to 0.25 Bohr$^{-1}$. \review{We selected Si in the diamond structure as example from the above benchmark, since for the $\omega=0.11$~Bohr$^-1$ used here, the deviations between PBE0$^\prime$ and PBE0 were the largest among the inorganic systems as shown in Fig.~\ref{fig:inorganics}.} We observe that the error in the gKS band gap changes almost linearly from -0.539~eV to -0.114~eV when going from $\omega=0.25$ to $0.08$~Bohr$^{-1}$. Below $\omega=0.08$~Bohr$^{-1}$ the error in the gKS band gap falls below 0.1 eV, and continues to converges towards the unmodified PBE0 result. A similar behavior is found for the error in the total energy, however the error seems to grow much faster beyond $\omega=0.15$~Bohr$^{-1}$. As a result of this analysis we find that the range-separation parameter $\omega$ controls the accuracy and the speed of the calculation at the same time. 

Overall, the use of the screening function even improves the results when comparing, e.g., predicted band gaps with the experimental values. For large band gap materials, the predicted band gaps for PBE0$^\prime$ remains close to the PBE0 results, but for smaller band gaps the predicted band gaps get significantly smaller, e.g., for Silicon the PBE0$^\prime$ band gap is 1.610~eV, while the PBE0 band gap is 1.823~eV and the actual zero-point corrected experimental value is around 1.23~eV~\cite{bludau1974temperature,engel2022zero}. In turn, it is also possible to interpret the proposed screening function as part of a new functional and construct the functional in a similar way as for the HSE model: Fixing the fraction of exact exchange to 0.25, the range-separation parameter can be optimized to resemble experimental band gaps. In the Silicon example, Figure~\ref{fig:errors}, the experimental band gap is recovered at $\omega = 0.25$ Bohr$^{-1}$. In this case, the speed-up factor would be even larger than for $\omega = 0.11$ Bohr$^{-1}$ as used before. 

\subsection{Organic crystals}

\begin{figure*}[t]
    \centering
    \includegraphics[width=\textwidth]{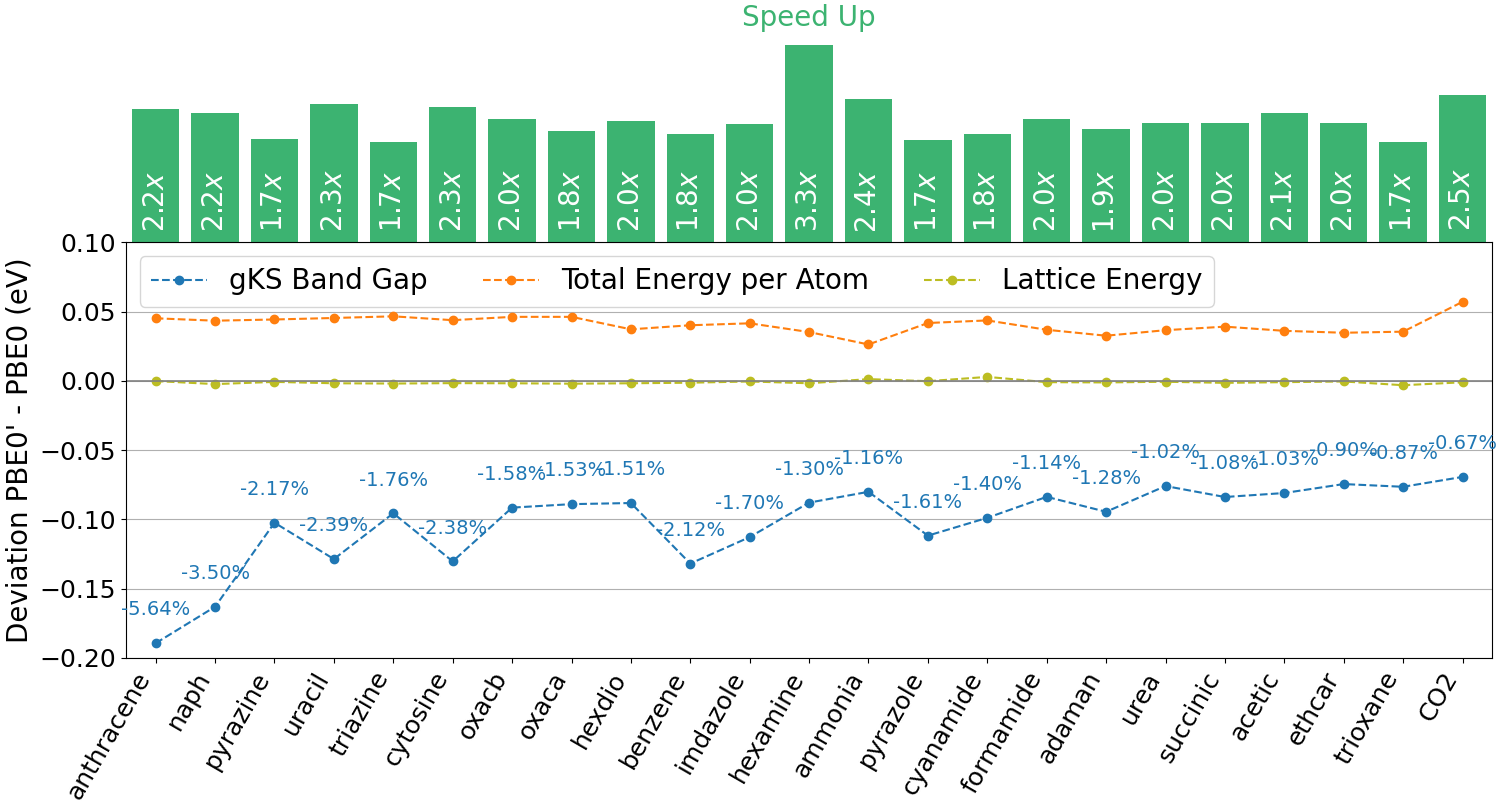}
    \caption{The deviations of the PBE0$^\prime$ functional from the unmodified PBE0 functional for the generalized Kohn-Sham (gKS) band gap (blue line), the total energy per atom (orange line), and the lattice energy $E_\text{latt}$ for the organic materials from the X23 data set. The materials are sorted from left to right with increasing PBE0 band gap size. The green bars give the speed-up for each material compared to a PBE0 calculation without EXX long-range approximation. The annotated numbers for the gKS band gaps are the relative errors w.r.t. to the original PBE0 gap.}
    \label{fig:x23}
\end{figure*}

We use the X23 dataset~\cite{reilly2013understanding} that consists of 23 organic crystals. The detailed listing defining chemical formula, space group, and abbreviations are given in the Supplemental Information, Table~2. We again use the PBE0 functional as test functional for our screening function and the many-body dispersion interaction with range-separated self-consistent screening MBD@rsSCS~\cite{ambrosetti2014,hermann2023libmbd} and the FHI-aims intermediate species defaults as well as the same, well-converged k-grid density as for the inorganic part. We perform a full SCF cycle including force evaluation that is representative of a single {\it ab initio} molecular dynamics step. The results are summarized in Figure~\ref{fig:x23} and all numerical details are listed in the Supplemental Information, Table~4. The organic crystals in Figure~\ref{fig:x23} are sorted again in ascending size of the gKS band gap. The errors in the band gaps behave similarly as for the inorganic materials: the error is the larger, the smaller the band gap. We find that the deviation in the total energy of the PBE0$^\prime$ functional from PBE0 is to large extent at the same level of accuracy for all tested organic crystals around 0.05 eV per atom. Also, the deviations in the computed lattice energy $E_\text{latt} = (E_\text{crystal} - N \cdot E_\text{molecule})/N$ are at most 3.1~meV and, thus, very low. The speed-up factors are around two and, thus, smaller then for the inorganic materials indicating that the density matrices associated with each system are already sparse, since the band gaps are large and there are typically no chemical bonds connecting the molecules in the organic crystal. 

\section{Conclusions}

We have introduced a screening of the Coulomb potential that is based on a first order Taylor expansion of the error function in terms of the HSE06 screening parameter. It captures a large part of the missing long-range Coulomb term of the PBE0 functional in an effective way - but at a much lower spatial extent of the interaction potential and therefore lower cost than the unmodified PBE0 function. In general, the screening function $s^\prime$ defined in ~\eqref{s_prime} can incorporated into all global hybrids, e.g., as explicitly tested in this work for PBE0, but also applicable in the B3LYP, M11, and many more. In principle, the long-range screening function defined in Eq.~\eqref{eq:lr_approx} can be used on its own for the long-range EXX contribution in long-range corrected hybrids, e.g. LC-$\omega$PBEh~\cite{vydrov2006assessment} and range-separated hybrid functionals. The screening function $s^\prime$, Eq.~\eqref{s_prime}, enables a significant computational speed-up as tested for inorganic and organic crystals when the EXX contributions are evaluated in real-space. The screening function $s^\prime$ has been implemented in FHI-aims~\cite{blum2009} as part of the localized resolution-of-identity framework RI-LVL~\cite{ihrig2015,levchenko2015,kokott2024efficient}. The range-separation parameter in the screening function controls the computational speed-up versus the accuracy in the long-range contribution. Large-scale simulations will benefit from a scaling effect, since less resources will be needed (most importantly in terms of memory), so the long-range Coulomb approximation is able to extend the limits of what can be currently computed.

Overall, the first-order approximated functional PBE0$^\prime$ is close to the full PBE0 functional, certainly within the errors typically expected between predictions from PBE0 compared to the experimental ground truth. Especially for larger scale simulations that require PBE0 for full accuracy, the PBE0$^\prime$ functional is an approximate, significantly faster alternative. The results presented in the present paper show significant promise of this approach for total energies, cohesive energies, lattice energies, and band gaps. We anticipate more detailed future tests of further observables (e.g. overall band structure, starting point quality for many-body perturbation theory, response properties derived from density functional perturbation theory etc) might be subject to similar speedups while essentially retaining full hybrid DFT precision. Although not tested in this work, the screening function should also lead to significant computational speed-up for larger non-periodic systems, e.g. like large molecules or cluster of molecules.

\review{
\section*{Supplementary Material}

The supplemental material contains a detailed listing defining stoichiometry, space group and abbreviations used for each system. We also list all absolute values for the deviations shown in Figs.~\ref{fig:inorganics} and \ref{fig:x23}.
}
\section*{Code and data availability}

The FHI-aims code is an academic community code and available to any academic group, including its source code, for a voluntary license fee, enabling, access to the full sources and development thereof by any academic research group. All data that supports this work is openly available in the NOMAD data base at \href{https://dx.doi.org/10.17172/NOMAD/2025.01.31-2}{10.17172/NOMAD/2025.01.31-2}.

\section*{Author declarations}

\subsection*{Conflict of Interest}

The authors have no conflicts to disclose.

\bibliographystyle{unsrt}
\bibliography{refs}

\newpage

\section*{Supplemental Information}

\begin{table*}[!htp]
    \centering
    \begin{tabular}{c|c|c}
        System Name & Fomula & Space Group No. \\
         Si& Si$_2$ & 227 \\
         GaAs & GaAs & 216 \\
         MAPI & H$_3$CNH$_3$PbI$_3$ & 1 \\
         CBTS & Ba$_3$Cu$_6$S$_{12}$Sn$_3$ & 144 \\
         AlAs & AlAs & 216 \\
         SiC & SiC & 216 \\
         ZnO & Zn$_2$O$_2$ & 186 \\
         Fe$_2$O$_3$ & Fe$_4$O$_6$ & 167 \\
         LiH & LiH & 225 \\
         Ga$_2$O$_3$ & Ga$_4$O$_6$ & 12 \\
         C & C$_2$ & 227 \\
         BN & BN & 216 \\
         Water & 64 $\cdot$ (H$_{2}$O) & 1 \\
         MgO & MgO & 225 \\
         SiO$_2$ & Si$_3$O$_6$ & 152 \\
         LiF & LiF & 225 \\
    \end{tabular}
    \caption{The name, the chemical formula, and the space group number of the benchmarked inorganic system.}
    \label{tab:list_inorganic}
\end{table*}

\begin{table*}[t]
    \centering
    \begin{tabular}{c|c|c|c}
        System Name & Full name & Fomula & Space Group No. \\
         anthracene & Anthracene & 2 $\cdot$ C$_{14}$H$_{10}$ & 14 \\
         naph & Naphthalene & 2 $\cdot$ C$_{10}$H$_{18}$ & 14 \\
         pyrazine & Pyrazine & 2 $\cdot$ C$_4$H$_4$N$_2$ & 58 \\
         uracil & Uracil & 4 $\cdot$ C$_4$H$_4$N$_2$O$_2$ & 14 \\
         triazine & s-Triazine & 6 $\cdot$ (HCN)$_3$ & 167 \\
         cytosine & Cytosine & 4 $\cdot$ C$_4$H$_5$N$_3$O & 19 \\
         oxacb & Oxalic acid ($\beta$) & 2 $\cdot$ (COOH)$_2$  & 14 \\
         oxaca & Oxalic acid ($\alpha$) & 4 $\cdot$ (COOH)$_2$ & 61 \\
         hexdio & 1,4-cyclohexanedione & 2 $\cdot$ (CH$_2$)$_4$(CO)$_2$ & 4 \\
         benzene & Benzene & 4 $\cdot$ C$_{6}$H$_{6}$ & 61  \\
         imdazole & Imidazole & 4 $\cdot$ (CH)$_3$(NH)N & 227 \\
         hexamine & Hexamine & (CH$_2$)$_6$N$_4$ & 217 \\
         ammonia & Ammonia & 4 $\cdot$ NH$_3$ & 198 \\
         pyrazole & Pyrazole & 8 $\cdot$ (CH)$_3$N$_2$H & 33 \\
         cyanamide & Cyanamide & 8 $\cdot$ CN$_2$H$_2$ & 61 \\
         formamide & Formamide & 4 $\cdot$ CH$_3$NO & 14 \\
         adaman& Adamantane & 2 $\cdot$ (CH)$_4$(CH$_2$)$_6$ & 114 \\
         urea& Urea & 2 $\cdot$ CO(NH$_2$)$_2$ & 113 \\
         succinic& Succinic acid & 2 $\cdot$ (CH$_2$)$_2$(CO$_2$H)$_2$ & 14 \\
         acetic& Acetic acid & 4 $\cdot$ CH$_3$COOH & 33 \\
         ethcar& Ethyl carbamate & 2 $\cdot$ CH$_3$CH$_2$OC(O)NH$_2$ & 2 \\
         trioxane& s-Trioxane & 6 $\cdot$ C$_3$H$_6$O$_3$ & 161 \\
         CO$_2$ & Carbon dioxide & 4 $\cdot$ CO$_2$ & 205 \\
    \end{tabular}
    \caption{The system and full name, the chemical formula, and the space group number of the benchmarked organic system.}
    \label{tab:list_organic}
\end{table*}

\begin{table*}[b]
    \centering
    \begin{tabular}{l|r|r|r|r|r|r}
        \hline
        \multicolumn{1}{c|}{Name} & \multicolumn{1}{c|}{$E_\text{tot}$/Atom} & \multicolumn{1}{c|}{$E_\text{coh}$/Atom} & gKS Band Gap & Nodes & CPUh & Speed \\
        \multicolumn{1}{c|}{(\#Atoms)} & \multicolumn{1}{c|}{(eV)} & \multicolumn{1}{c|}{(eV)} &\multicolumn{1}{c|}{(eV)} &  & & Up\\\hline
        AlAs         & -34255.385 (+0.145) & -3.738 (-0.004) & 2.544 (-0.177) &   1 (1) &   2.26 &   3.0 \\
        BN           &  -1084.688 (+0.054) & -6.849 (-0.013) & 6.511 (-0.116) &   1 (1) &   9.93 &   3.0 \\
        C      &  -1037.015 (+0.059) & -7.632 (-0.007) & 5.963 (-0.131) &   1 (1) &   7.93 &   2.9 \\
        CBTS   & -66192.311 (+0.155) & -3.802 (-0.013) & 2.391 (-0.146) &   1 (2) &  39.14 &   3.5 \\
        Fe$_2$O$_3$       & -15135.842 (+0.088) & -4.717 (-0.015) & 4.009 (-0.092) &   1 (4) &  41.69 &  3.9 \\
        Ga$_2$O$_3$       & -22502.861 (+0.079) & -4.635 (-0.017) & 4.756 (-0.108) &   1 (2) &  20.16 &   3.2 \\
        GaAs        & -57539.298 (+0.164) & -3.177 ( 0.004) & 1.784 (-0.162) &   1 (1) &   2.19 &   3.7 \\
        LiF          &  -1463.952 (+0.036) & -4.254 (-0.030) & 11.689 (-0.059) &   1 (1) &   3.99 &   2.4 \\
        LiH          &   -110.755 (+0.030) &  -2.384 (-0.022) & 4.549 (-0.133) &   1 (1) &   3.81 &   2.6 \\
        MAPI        & -98547.973 (+0.099) & -3.113 ( 0.003) &  2.209 (-0.200) &   1 (1) &   4.45 &   3.0 \\
        MgO          &  -3754.184 (+0.058) & -4.934 (-0.038) &  7.123 (-0.101) &   1 (1) &   5.16 &   2.1 \\
        Si (2)          &  -7901.864 (+0.134) & -4.574 ( 0.000) & 1.610 (-0.193) &   1 (1) &   2.02 &   3.8 \\
        SiC          &  -4469.735 (+0.087) & -6.340 (-0.014)&  2.878 (-0.151) &   1 (1) &   4.94 &   3.0 \\
        SiO$_2$        &  -4001.103 (+0.061) & -6.176 (-0.018) &  8.590 (-0.077) &   1 (1) &   7.20 &   2.7 \\
        Water      &   -693.709 (+0.028) & -3.377 (-0.002) & 6.815 (-0.070) &   2 (5) &  28.20 &   2.1 \\
        ZnO          & -25583.782 (+0.084) & -3.501 (-0.020) &3.075 (-0.108) &   1 (1) &   8.29 &   3.7 \\ \hline
    \end{tabular}
    \caption{The dataset of 16 inorganic crystals computed with the PBE0$^\prime$ functional with an exchange mixing parameter $\alpha=0.25$ and a range separation parameter $\omega=0.11$~Bohr$^{-1}$. All calculations use the intermediate species defaults from the FHI-aims software package and a k-point density in each reciprocal direction with at least 6 points per \AA$^-1$. E$_\text{tot}$/atom: total energy per atom; the deviation to the unmodified PBE0 functional is given in parentheses. gKS Band Gap: The Kohn-Sham band gap; the deviation to the unmodified PBE0 functional is given in parentheses. $\Delta F_\text{max}$: The maximum deviation of a force component w.r.t. to the unmodified PBE0 functional. Nodes: Number of nodes; minimum number of nodes need for the undmodified PBE0 functional are given in parentheses. CPUh: CPU hours computed on nodes with 72 CPUs per node. Speed-up: The speed-up factor for the PBE0$^\prime$ functional compared to the unmodified PBE0 functional computed as the ratio of CPU hours. Most of the systems only contain a small number of atoms per unit cell and, thus, most of them could be run even with less computational resources than a full node as shown here.}
    \label{tab:inorganics}
\end{table*}

\begin{table*}[b]
    \centering
    \begin{tabular}{l|r|r|r|r|r|r|r}
        \hline
        \multicolumn{1}{c|}{Name}  & \multicolumn{1}{c|}{E$_\text{tot}$/Atom}  & \multicolumn{1}{c|}{$E_\text{lattice}$}   & \multicolumn{1}{c|}{gKS Band Gap}    & $\Delta F_\text{max}$& Nodes & CPUh & speed \\
        \multicolumn{1}{c|}{(\#Atoms)} & \multicolumn{1}{c|}{(eV)} & \multicolumn{1}{c|}{(eV)} & \multicolumn{1}{c|}{(eV)} & eV/A & & & up\\\hline
        CO$_2$        & -1711.352 (+0.057) & -0.2836 (-0.0010)& 10.292 (-0.069) &  0.040 &   1 (2) &   3.64 &   2.5 \\
        acetic      &  -779.550 (+0.036) & -0.8240 (-0.0009) &  7.787 (-0.081) &  0.054 &   1 (3) &   9.53 &   2.1 \\
        adaman      &  -408.883 (+0.033) & -0.8542 (-0.0011) &  7.298 (-0.094) &  0.025 &   2 (6) &  17.79 &   1.9 \\
        ammonia     &  -384.900 (+0.026) & -0.4372 ( 0.0013) &  6.804 (-0.080) &  0.025 &   1 (2) &   2.92 &   2.4 \\
        anthracene &  -611.646 (+0.045) & -1.2296 (-0.0001) &  3.166 (-0.189) &  0.035 &   2 (8) &  25.54 &   2.2 \\
        benzene     &  -526.574 (+0.040) & -0.6157 (-0.0013) &  6.079 (-0.132) &  0.027 &   2 (5) &  16.71 &   1.8 \\
        cyanamide   &  -809.960 (+0.044) & -0.9466 ( 0.0029) &  6.964 (-0.099) &  0.092 &   2 (4) &  17.66 &   1.8 \\
        cytosine    &  -826.902 (+0.044) & -1.7717 (-0.0016) &  5.351 (-0.130) &  0.050 &   2 (6) &  26.94 &   2.3 \\
        ethcar      &  -677.906 (+0.035) & -0.9777 (-0.0004) &  8.165 (-0.074) &  0.051 &   1 (3) &   8.789 &   2.0 \\
        formamide   &  -770.849 (+0.037) & -0.8783 (-0.0009) &  7.278 (-0.084) &  0.047 &   1 (3) &   9.12 &   2.0 \\
        hexamine    &  -562.603 (+0.035) & -0.9886 (-0.0017) &  6.687 (-0.088) &  0.019 &   1 (3) &   9.70 &   3.3 \\
        hexdio      &  -653.025 (+0.037) & -1.0109 (-0.0017) &  5.764 (-0.088) &  0.054 &   1 (3) &  11.48 &   2.0 \\
        imdazole    &  -684.062 (+0.042) & -1.0030 (-0.0004) &  6.514 (-0.113) &  0.042 &   2 (4) &  15.41 &   2.0 \\
        naph        &  -583.294 (+0.043) & -0.9410 (-0.0023) &  4.488 (-0.163) &  0.030 &   1 (4) &  14.19 &   2.2 \\
        oxaca       & -1287.544 (+0.046) & -1.1177 (-0.0020) &  5.725 (-0.089) &  0.046 &   1 (4) &  15.74 &   1.8 \\
        oxacb       & -1287.542 (+0.046) & -1.1019 (-0.0017) &  5.683 (-0.091) &  0.046 &   1 (2) &   8.39 &   2.0 \\
        pyrazine    &  -719.254 (+0.044) & -0.7189 (-0.0007) &  4.615 (-0.102) &  0.047 &   1 (2) &  10.24 &   1.7 \\
        pyrazole    &  -684.000 (+0.042) & -0.8845 (-0.0002) &  6.815 (-0.112) &  0.047 &   2 (7) &  31.70 &   1.7 \\
        succinic    &  -888.585 (+0.039) & -1.4514 (-0.0014) &  7.675 (-0.084) &  0.047 &   1 (3) &  11.96 &   2.0 \\
        triazine    &  -847.744 (+0.047) & -0.6612 (-0.0019) &  5.334 (-0.096) &  0.048 &   2 (7) &  33.00 &   1.7 \\
        trioxane    &  -779.364 (+0.036) & -0.7132 (-0.0031) &  8.758 (-0.076) &  0.032 &   2 (7) &  32.90 &   1.7 \\
        uracil      &  -940.959 (+0.045) & -1.5238 (-0.0017) &  5.253 (-0.129) &  0.056 &   2 (7) &  26.99 &   2.3 \\
        urea       &  -766.581 (+0.037) & -1.1843 (-0.0006) &  7.377 (-0.076) &  0.051 &   1 (2) &   5.17 &   2.0 \\ \hline
    \end{tabular}
    \caption{The X23 dataset of organic crystals computed with the PBE0$^\prime$ functional with an exchange mixing parameter $\alpha=0.25$ and a range separation parameter $\omega=0.11$~Bohr$^{-1}$. All calculations use the intermediate species defaults and a k-point density in each reciprocal direction $>$6/$AA^{-1}$. E$_\text{tot}$/atom: total energy per atom; the deviation to the unmodified PBE0 functional is given in parentheses. $E_\text{lattice}$: The lattice energy per molecule; the deviation to the unmodified PBE0 functional is given in parentheses. gKS Band Gap: The Kohn-Sham band gap; the deviation to the unmodified PBE0 functional is given in parentheses. $\Delta F_\text{max}$: The maximum deviation of a force component w.r.t. to the unmodified PBE0 functional. Nodes: Number of nodes; minimum number of nodes need for the undmodified PBE0 functional are given in parentheses. CPUh: CPU hours computed on nodes with 72 CPUs per node. Speed-up: The speed-up factor for the PBE0$^\prime$ functional compared to the unmodified PBE0 functional computed as the ratio of CPU hours.}
    \label{tab:X23}
\end{table*}

\end{document}